\documentclass[aps,preprint]{revtex4}

\usepackage{graphicx}
\begin{document}

\title{Storage Capacity of Two-dimensional Neural Networks}
\author{Shinsuke Koyama} 
\email{s_koyama@complex.eng.hokudai.ac.jp}
\affiliation{Complex Systems Engineering, Graduate School of
Engineering, Hokkaido University, 
N13-W8, Kita-ku, Sapporo 8628, Japan}
\date{\today}

\begin{abstract}
We investigate the maximum number of embedded patterns in the 
two-dimensional Hopfield model.
The grand state energies of two specific network states, namely, 
the energies of the pure-ferromagnetic state and the state of 
specific one stored pattern are calculated exactly in terms of 
the correlation function of the ferromagnetic Ising model.
We also investigate the energy landscape around them by computer 
simulations.
Taking into account the qualitative features of the phase diagrams 
obtained by Nishimori, Whyte and Sherrington 
[Phys. Rev. E {\bf 51}, 3628 (1995)], 
we conclude that the network cannot retrieve more than three patterns.
\end{abstract}

\maketitle

\section{Introduction}
The Hopfield model \cite{bib:Hopfield} is one of the simplest
mathematical models which explains associative memory.
This model is characterized by binary state neurons and each neuron 
is represented by Ising spin.
In this model system, arbitrary two neurons are interacted each other 
{\it via} so called Hebb rule.
The Hebb rule is one of the standard learning rules of the patterns 
$\xi_i^{\mu}\ (i=1,\ldots ,N;\ \mu=1,\ldots ,p)$ and determines the 
strength of the interaction $J_{ij}$ between the $i$ and $j$-th neurons, 
say, $S_i$ and $S_j$, as
\begin{equation}
 J_{ij} = \frac{1}{N}\sum_{\mu}\xi_i^{\mu}\xi_j^{\mu}  ,
\end{equation}
where $\mu$ means the number of embedded patterns and $N$ denotes the 
number of neurons.
These features have been investigated by statistical mechanics deeply 
\cite{bib:Amit,bib:intro_nc}.
Actually, up to now, various extensions and generalizations of the 
Hopfield model were proposed and these properties were investigated 
from statistical mechanical point of view 
(see, for example \cite{bib:intro_nc}).
However, little is known about the properties of the Hopfield model 
in which the length of interactions $J_{ij}$ is restricted to the 
nearest neighboring neurons.
By the analogy to the spin system in statistical mechanics, we call 
this type of the Hopfield model \textit{finite-dimensional Hopfield model}.
Inspired by the study of Nishimori, Whyte and Sherrington
\cite{bib:Nishimori}, in this paper, we consider 
the finite-dimensional Hopfield model storing structured patterns.
In general, it is hard to analyze such finite-dimensional systems
explicitly.
However, one can derive several rigorous results of thermodynamic 
properties of the system by assistant of the Pierls arguments and 
the gauge transformations \cite{bib:Nishimori}.
Although the qualitative features of the phase diagrams of the system 
became clear by these analyses,
nobody yet succeeded in deriving their quantitative behavior at all.
In this paper, we analyze the storage capacity of the system
quantitatively.
We restrict ourselves to the case of two-dimensional system on the 
square lattice.

This paper is organized as follows. 
In Sec.~\ref{sec:definition}, we explain our model system.
In Sec.~\ref{sec:phase_diag}, we briefly review the qualitative features
of the phase diagram obtained by 
Nishimori {\it et al} \cite{bib:Nishimori}.
In Sec.~\ref{sec:analysis}, we analyze the storage capacity of our 
model system.
Section \ref{sec:conclusion} is devoted to discussion of all results we
obtain.

\section{Definition of the system} \label{sec:definition}
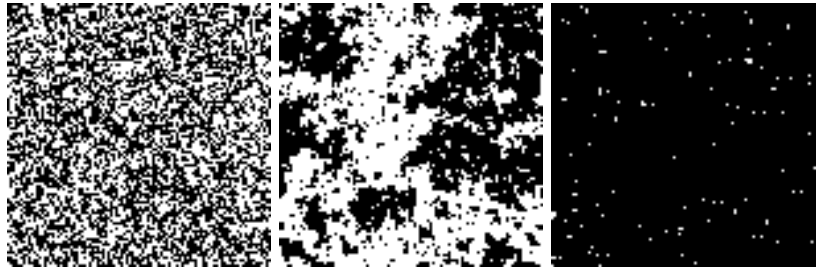
\begin{figure}[htbp]

 \caption{Typical examples of embedded patterns. 
 The size of patterns is $100 \times 100$. 
From the left to the right, the values of parameter $J_0/\sqrt{p}$ are 
 $0.10$, $0.42$ and $0.60$.}
 \label{img:snap_shot}
\end{figure}

In this section, we define our model systems.
The Hamiltonian of the system is given as
\begin{equation}
 H = -\sum_{\langle ij \rangle} J_{ij} S_i S_j  , \label{eq:hamiltonian}
\end{equation}
where $S_i(i=1,\ldots ,N)$ are the states of the neuron taking binary 
value $\pm 1$, and $J_{ij}$ is the strength of the interaction between 
$S_i$ and $S_j$.
The summation $\sum_{\langle ij \rangle}$ appearing in 
Eq.~(\ref{eq:hamiltonian}) runs
over nearest neighboring neurons on a square lattice.
We chose the {\it short range} Hebb rule as a interaction which is 
given by
\begin{equation}
 J_{ij} = \frac{1}{\sqrt{p}}\sum_{\mu=1}^{p} 
  \xi_i^{\mu}\xi_j^{\mu}   \label{eq:Hebb_rule}
\end{equation}
for the nearest neighboring sites $\langle ij \rangle$ and $J_{ij}=0$ 
otherwise.
Here, $p$ is the number of embedded patterns and 
$\xi_i^{\mu}=\pm 1 \ (\mu=1,\ldots ,p;\ i=1,\ldots ,N)$. 
Let us consider about the probability distribution of a set of 
patterns 
$\{ \xi_i^{\mu} \} ( \equiv 
\{\xi_i^{\mu} \  | \ i=1,\dots ,N; \mu=1,\ldots ,p\} )$.
We suppose that the probability that arbitrary nearest neighboring sites 
of $\mu$-th pattern are $\xi_i^{\mu}$ and $\xi_j^{\mu}$ respectively
is proportional to 
\begin{equation}
 \exp\left( \frac{J_0}{\sqrt{p}}\xi_i^{\mu}\xi_j^{\mu} \right)  , 
  \label{eq:prob_nearest}
\end{equation} 
where the parameter $J_0$ controls the degree of the correlation between 
arbitrary nearest neighboring sites.
Let us write $P(\xi_i^{\mu} = \xi_j^{\mu})$ and 
$P(\xi_i^{\mu} = -\xi_j^{\mu})$ 
as the  probability of $\xi_i^{\mu} = \xi_j^{\mu}$ and 
$\xi_i^{\mu} = -\xi_j^{\mu}$, respectively.
Then, the ratio of the former to the later is given by 
\begin{equation}
 \frac{P(\xi_i^{\mu} = \xi_j^{\mu})}{P(\xi_i^{\mu} = -\xi_j^{\mu})}
 = \exp\left(\frac{2J_0}{\sqrt{p}}\right) . \label{eq:ratio_prob}
\end{equation}
For the case of $J_0=0$, Eq.~$(\ref{eq:ratio_prob})$ becomes 
$P(\xi_i^{\mu} = \xi_j^{\mu})=P(\xi_i^{\mu} = -\xi_j^{\mu})$.
Hence there is no correlation between $i$ and $j$ sites of $\mu$-th 
patterns and embedded patterns correspond to ``random patterns''.
On the other hand, for the case of $J_0 \to \infty$, 
we obtain  $P(\xi_i^{\mu}=-\xi_j^{\mu})=0$. 
Namely, the value of $\xi_i^{\mu}$ is same as that of $\xi_j^{\mu}$ with
probability one.
Applying Eq.~$(\ref{eq:prob_nearest})$ to all nearest neighboring pairs 
$\langle ij \rangle$, 
we obtain the probability distribution for $\mu$-th pattern as 
\begin{equation}
 \frac{1}{Z_0\left( \frac{J_0}{\sqrt{p}} \right)}
  \exp\left(\frac{J_0}{\sqrt{p}}\sum_{\langle ij \rangle} 
       \xi_i^{\mu}\xi_j^{\mu}\right)  ,
  \label{eq:prob_one_pattern}
\end{equation}
where $Z_0(J_0/\sqrt{p})$ is the normalization factor given by
\begin{equation}
 Z_0\left( \frac{J_0}{\sqrt{p}}\right) =
  \sum_{ \{ \xi_i \} } 
  \exp \left( \frac{J_0}{\sqrt{p}} 
	\sum_{\langle ij \rangle} \xi_i \xi_j \right)  .
  \label{eq:partition}
\end{equation}
Supposing that each pattern is generated by
Eq.~$(\ref{eq:prob_one_pattern})$ independently,
a set of embedded patterns $\{\xi_i^{\mu}\}$ 
is generated by the following probability distribution
\begin{equation}
 P(\{ \xi_i^{\mu}\}) = c \prod_{\mu=1}^p
  \exp \left( \frac{J_0}{\sqrt{p}} \sum_{\langle ij \rangle}
	\xi_i^{\mu} \xi_j^{\mu} \right) .
  \label{eq:distribution}
\end{equation}
Here, the normalization factor $c$ is given by $\{
Z_0(J_0/\sqrt{p})\}^p$.
Note that Eq~$(\ref{eq:prob_one_pattern})$ corresponds to 
the Boltzmann weight of the two-dimensional ferromagnetic Ising
model on the square lattice whose interaction is given by $1/\sqrt{p}$ at 
temperature $1/J_0$.
Hence embedded patterns are same as snapshots of 
equilibrium Monte Carlo simulations for it.
This Ising model was explicitly solved in \cite{bib:Onsager} and it is 
known that there is a critical point at $K=K_c=0.44$, where $K$ denotes 
the ratio of the temperature to the strength of the interaction.
This model has a ferromagnetic solution for $J_0/\sqrt{p} > K_c$ and 
a paramagnetic solution for $J_0/\sqrt{p} < K_c$.
From this fact, in our model system, embedded patterns have a long-range 
order for $J_0 > K_c\sqrt{p}$, 
on the other hand, there is no long-range correlation for 
$J_0 < K_c\sqrt{p}$.
Fig.~\ref{img:snap_shot} shows typical three examples of embedded patterns.

In the next section, we briefly review the features of the phase
diagrams of our model systems obtained by Nishimori 
{\it et al} \cite{bib:Nishimori}.

\section{Generic qualitative phase diagram} \label{sec:phase_diag}
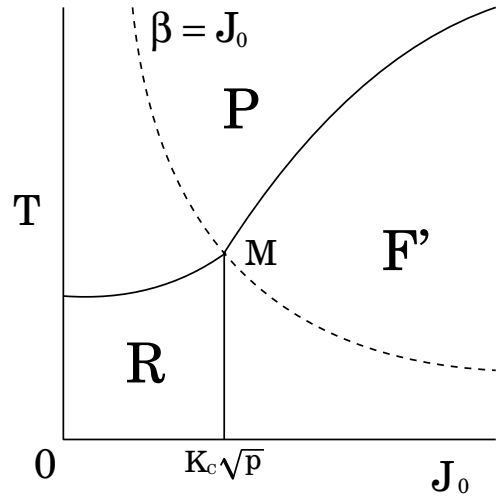
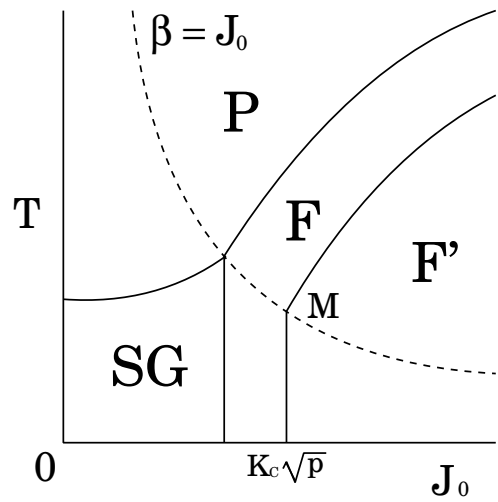
\begin{figure}[htbp]

  \caption{The qualitative phase diagram with axes of temperature $T$ 
  and a parameter $J_0$ controlling the structure of patterns.
  The value of $p$ is fixed.
(a) For small $p$, there are three phases: paramagnetic P, retrieval R, 
and ferromagnetic with finite overlap F'.
(b) For large $p$, there appears a ferromagnetic phase without retrieval
  order F, and the retrieval phase in the small-$p$ case is replaced
  by the spin glass phase.
}
  \label{img:phase}
\end{figure}

Before we explain our analysis of maximum number of embedded patterns,
in this section, 
we briefly review the results by Nishimori {\it et al} \cite{bib:Nishimori}.
Note that their treatments, namely, the gauge transformations and the 
Peierls arguments are applied to not only the two-dimensional 
systems but also the systems in arbitrary dimension.
Fig.~\ref{img:phase} shows the qualitative phase diagram with 
axes of temperature $T$ and a parameter $J_0$ controlling the structure of
patterns for a fixed value of $p$.
In general, this system has three phases:
paramagnetic(P), ferromagnetic(F), 
and retrieval(R)(or spin glass(SG)) phases.
Each phase is characterized by the following three order parameters 
\begin{eqnarray}
 q &=&  \frac{1}{N}\sum_{i=1}^N\langle S_i \rangle^2  , \\
 m_F &=& \frac{1}{N}\sum_{i=1}^N\langle S_i \rangle  , \\
 m_R &=& \frac{1}{N}\sum_{i=1}^N\xi_i^{\mu}\langle S_i \rangle ,
\end{eqnarray}
where $\langle \cdots \rangle$ means thermodynamic average.
The above three order parameters $q$, $m_F$ and $m_R$ represent 
the Edwards-Anderson spin glass order parameter, the ferromagnetic 
order parameter, and the overlap between the $\mu$-th pattern 
 and the network state ${\{S_i\}}$, respectively.
The paramagnetic, ferromagnetic, retrieval and spin grass phase are 
characterized by the above three order parameters as 
$q=m_F=m_R=0;$ 
$q>0, m_F>0; $ 
$q>0, m_R>0$ and 
$q>0, m_F=m_R=0$, respectively.

Applying the gauge transformations to this system,
we obtain the internal energy $[\langle E \rangle]$ and 
the overlap $[m_R]$ on the lines $\beta(=1/T)=J_0$ in the phase diagram
as follows.
\begin{equation}
 [\langle E \rangle] = pE_0\left(\frac{\beta}{\sqrt{p}}\right),
  \label{eq:internal_energy}
\end{equation}
\begin{equation}
 [m_R]= m_0\left(\frac{\beta}{\sqrt{p}}\right),
  \label{eq:overlap_mag}
\end{equation}
where $E_0(\beta/\sqrt{p})$ is the internal energy of the ferromagnetic 
Ising model corresponding to the partition function
$Z_0(\beta/\sqrt{p})$,
and $m_0(\beta/\sqrt{p})$ is the spontaneous magnetization of the same
model.
Here, the expression of $[\cdots]$ means the average over 
the distribution~(\ref{eq:distribution}).
Both $E_0(\beta/\sqrt{p})$ and $m_0(\beta/\sqrt{p})$ 
generally have singularities 
at some critical point when spatial dimensionality exceeds one.
Let us suppose that $\beta/\sqrt{p}=K_c$ is the critical point.
Then Eqs~$(\ref{eq:internal_energy})$ and $(\ref{eq:overlap_mag})$ 
mean that the internal energy and the overlap of our system 
have the same singularity at $\beta/\sqrt{p}=K_c$.
This implies a phase transition and 
the boundary between $[m_R] \neq 0$ and $[m_R] = 0$ crosses
this point on the line $\beta(=1/T)=J_0$
(it is denoted by M in Fig.~\ref{img:phase}).
We should notice that this point is also the critical point of the 
embedded patterns.

When the value of $p$ is small, 
a typical phase diagram is given by Fig.~\ref{img:phase}(a).
For this case, there is a possibility to exist the retrieval phase in the 
region denoted by R. 
In the same figure, 
The region F' is the ferromagnetic phase with finite overlap.
It is important to notice  that this system is not meaningful as an 
associative memory
for $J_0 > K_c\sqrt{p}$ since embedded patterns have a long-range 
ferromagnetic order like the right of Fig.~\ref{img:snap_shot},
and patterns become correlated each other.
With this fact in mind, we do not regard this region as a retrieval phase.
When the value of $p$ increases, 
the critical point $J_0=K_c\sqrt{p}$ moves to right.
On the other hand, by using the Peierls argument,
the ferromagnetic phase still exists in the almost same region.
Taking those facts into account, 
there exists a critical number of patterns $p_c$ 
above which $(p>p_c)$ the ferromagnetic phase is split into two regions,
that is, the ferromagnetic phase with finite overlap (F') and the 
ferromagnetic phase without retrieval order (F).
At the same time, the retrieval phase in the small-$p$ case is replaced 
by the spin glass phase because the region with finite overlap is limited 
to F' (Fig.~\ref{img:phase}(b)).
In summary,
there exists a critical number of patterns $p_c$ blow which $(p<p_c)$ 
the retrieval phase exists there if any, while for $p>p_c$ the 
retrieval phase vanishes and this network cannot retrieve embedded patterns.

However, the value of $p_c$ is not yet evaluated quantitatively at all
in \cite{bib:Nishimori}.
Following section, we investigate it especially in the case of 
two-dimensional system on the square lattice.

\section{Analysis of the system} \label{sec:analysis}
\subsection{The case of $p=1$}
We begin with the case of $p=1$.
In this case, the system is identical to the ferromagnetic Ising model 
and the retrieval solution of the system corresponds to the 
ferromagnetic solution of the ferromagnetic Ising model.
In the case of square lattice, the critical temperature of the 
ferromagnetic Ising model is $T_c=2.27$,
therefore the system has a retrieval solution at $T<T_c$.

\subsection{The case of $p \ge 3$}
We next analyze the case of $p\ge 3$.
There is a good evidence to show that the retrieval phase does not exist 
for this case. 
Let us start with investigating if there is a state
which has a smaller energy than retrieval one.
Substituting $S_i=\xi_i^1 (\textrm{for all}\ i)$ into
Eq.~(\ref{eq:hamiltonian}) and averaging it over the 
distribution~(\ref{eq:distribution}), 
we obtain the energy per neuron of the pure retrieval state
\begin{eqnarray}
 [E_R] &=& \frac{1}{N} 
  \left[ -\frac{1}{\sqrt{p}}\sum_{\langle ij \rangle}\sum_{\mu=1}^p
   \xi_i^{\mu}\xi_j^{\mu}\xi_i^1\xi_j^1
 \right] \nonumber\\
 &=&
  -\frac{2}{\sqrt{p}}
  \left\{ 1+(p-1)C_1\left( \frac{J_0}{\sqrt{p}}\right)^2 \right\} ,
  \label{eq:energy_R}
\end{eqnarray}
where $C_1(J_0/\sqrt{p})$ is the nearest-neighbor correlation function
of the ferromagnetic Ising model on the square lattice.
The explicit form of $C_1$ is 
\begin{equation}
 C_1\left( \frac{J_0}{\sqrt{p}} \right) =
  \frac{1}{Z_0\left(\frac{J_0}{\sqrt{p}}\right)}
  \sum_{ \{ \xi_i \} } \xi_i\xi_j
\exp\left( \frac{J_0}{\sqrt{p}}\sum_{\langle ij \rangle}\xi_i\xi_j\right) .
\label{eq:correlation}
\end{equation}
We also rewrite the energy per neuron of the pure ferromagnetic 
state in terms of $C_1(J_0/\sqrt{p})$.
Substituting $S_i=1 (\textrm{for all}\ i)$ into Eq.~(\ref{eq:hamiltonian}) 
and averaging it over the distribution~(\ref{eq:distribution}), we obtain
\begin{eqnarray}
 [E_F] &=& \frac{1}{N}
  \left[
   -\frac{1}{\sqrt{p}}\sum_{\langle ij \rangle}\sum_{\mu=1}^p
   \xi_i^{\mu}\xi_j^{\mu}
 \right] \nonumber \\
 &=&
 -2\sqrt{p}C_1\left(\frac{J_0}{\sqrt{p}}\right) .
  \label{eq:energy_F}
\end{eqnarray}

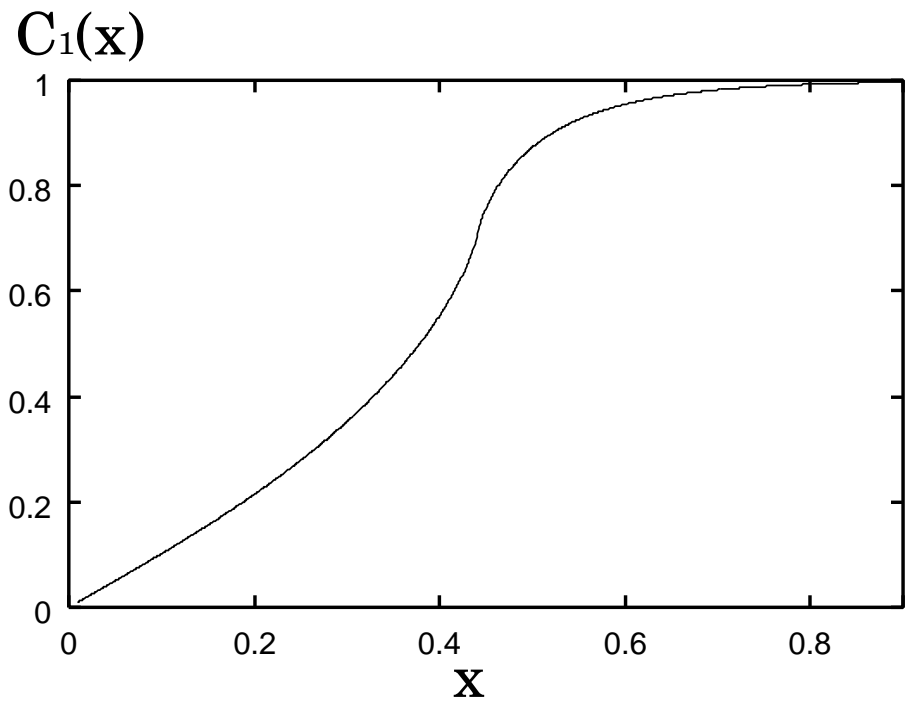
\begin{figure}[htbp]

  \caption{The shape of the correlation function $C_1(x)$ .}
  \label{img:correlation}
\end{figure}

We next investigate the properties of the function $C_1(J_0/\sqrt{p})$.
It is written in terms of the partition function~(\ref{eq:partition}) 
as follows.
\begin{equation}
  C_1\left( \frac{J_0}{\sqrt{p}} \right) =
   \frac{1}{2N}
   \frac{\partial \log Z_0(J_0/\sqrt{p})}{\partial (J_0/\sqrt{p})}
   \label{eq:corr}
\end{equation}
It is important to bear in mind that  $\log Z_0(J_0/\sqrt{p})$ is 
explicitly solved in \cite{bib:Onsager} as follows.
\begin{equation}
 \frac{1}{N}\log Z_0\left( \frac{J_0}{\sqrt{p}} \right) 
  = 
  \log \left( 2\cosh \frac{2J_0}{\sqrt{p}} \right) +
  \frac{1}{2\pi^2}
  \int_0^{\pi}\int_0^{\pi} \log (1-4\kappa\cos\omega_1\omega_2)
  d\omega_1 d\omega_2, \label{eq:onsager}
\end{equation}
where 
\begin{equation}
 2\kappa = \frac{\tanh (2J_0/\sqrt{p})}{\cosh (2J_0/\sqrt{p})} \ .
\end{equation}
Substituting 
$2\cosh\mu = 1/2\kappa|\cos\omega_1 |$ into Eq.~$(\ref{eq:onsager})$ 
and using
\begin{equation}
 \int_0^{2\pi}\log(2\cosh\mu-2\cos\omega)d\omega = 2\pi\mu,
\end{equation}
\begin{equation}
 \mu = \cosh^{-1}y = \log \left( y+\sqrt{y^2-1} \right),\quad 
  y = \frac{1}{4\kappa | \cos\omega_1 |} \  ,
\end{equation}
we obtain
\begin{equation}
\frac{1}{N}\log Z_0\left( \frac{J_0}{\sqrt{p}} \right) 
 = \log\left( 2\cosh \frac{2J_0}{\sqrt{p}} \right)
 + \frac{1}{2\pi}\int_0^{\pi} \log\frac{1}{2}
  \left(1+\sqrt{1-(4\kappa)^2\sin^2 \varphi} \right) d\varphi .
  \label{eq:onsager_r}
\end{equation}
Substituting Eq.~$(\ref{eq:onsager_r})$ into the right hand side of 
Eq.~(\ref{eq:corr}), 
explicit solution of $C_1(J_0/\sqrt{p})$ is written by
\begin{equation}
 C_1\left(\frac{J_0}{\sqrt{p}}\right) = \frac{1}{2}
\coth \frac{2J_0}{\sqrt{p}} \cdot 
\left( 1+ \frac{2}{\pi} \kappa_1'L \right) ,
\label{eq:explicit_corr}
\end{equation}
where $L$ is the complete elliptic integral, namely,  
\begin{equation}
 L = \int_0^{\pi/2} \frac{d\varphi}{\sqrt{1-\kappa_1^2 \sin^2\varphi}} \ ,
\end{equation}
with $\kappa_1 = 4\kappa$ and
\begin{equation}
 \kappa_1' = 2\tanh^2 \frac{2J_0}{\sqrt{p}} -1 .
\end{equation}
Fig.~\ref{img:correlation} shows the shape of 
Eq.~(\ref{eq:explicit_corr}).
From Eqs.~(\ref{eq:energy_R}), (\ref{eq:energy_F}) and
(\ref{eq:explicit_corr}), we obtain rigorous values of 
grand state energies of the pure retrieval and the pure ferromagnetic 
states.

Now we calculate the condition that the pure ferromagnetic state has 
a smaller energy than the retrieval one, 
that is to say, 
the condition for $[E_R] > [E_F]$.
We rewrite this inequality by using Eqs.~(\ref{eq:energy_R}) 
and (\ref{eq:energy_F}), 
then we obtain
\begin{equation}
 \frac{1}{p-1} < C_1\left(\frac{J_0}{\sqrt{p}}\right) < 1 . 
  \label{eq:condition_1}
\end{equation} 
As $C_1^{-1}$ is a monotonically increasing function and 
$C_1^{-1}(1)=\infty$ (see Fig .\ref{img:correlation}),
Eq.~(\ref{eq:condition_1}) leads to 
\begin{equation}
 \frac{J_0}{\sqrt{p}} >  C_1^{-1}\left(\frac{1}{p-1}\right) \equiv U ,
 \label{eq:condition_2}
\end{equation}
where $C_1^{-1}$ denotes the inverse of the function $C_1$. 
We find that the pure ferromagnetic state has a smaller energy than 
the pure retrieval one as long as this condition is satisfied.
As $J_0$ is finite, $p \ge 3$ must be satisfied
because $C_1^{-1}$ is a monotonically increasing function and 
$C_1^{-1}(1)=\infty$.
Further, it is easy to confirm the following inequality.
\begin{equation}
 \frac{\partial([E_F]-[E_R])}{\partial p} < 0
\end{equation}
Taking into account $[E_F]-[E_R] < 0$ under the condition 
$(\ref{eq:condition_2})$ and $p \ge 3$,
$[E_F]$ gets smaller relatively as $p$ gets larger.
As the result,
the pure ferromagnetic state has a smaller energy than the pure 
retrieval one for $T=0$, $J_0>U\sqrt{p}$ and $p \ge 3$.
The value of $U$ is $0.38$ for $p=3$, and $U < 0.38$ for $p>3$
because of the monotonically increasing property of $C_1^{-1}$.
Note that $U$ is always less than $K_c=0.44$.

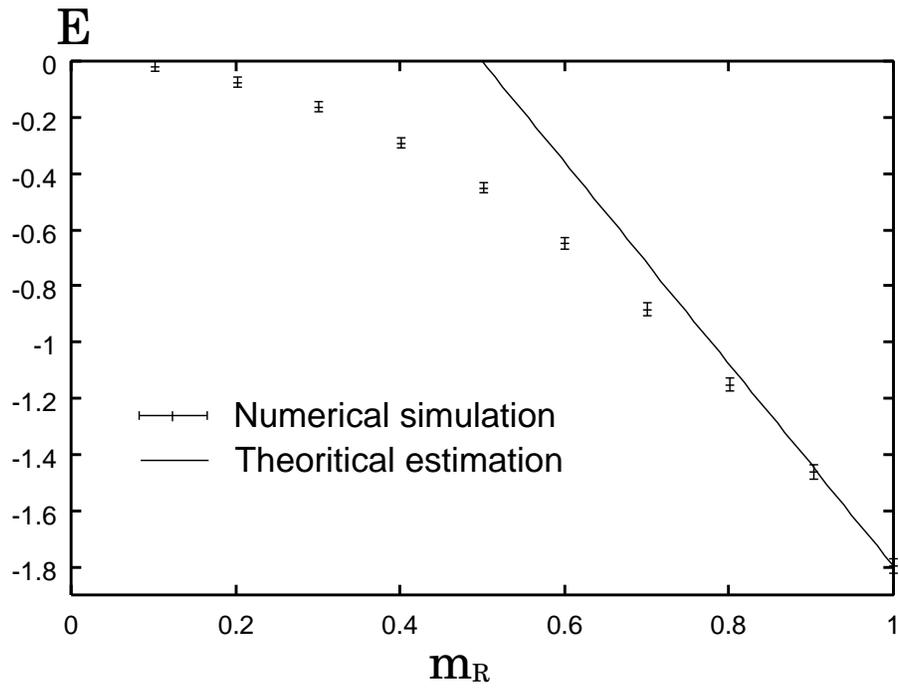
\begin{figure}[htbp]

  \caption{The average energies $E$ of the retrieval states with 
   overlap $m_R$.
  The parameter values are $p=3$ and $J_0\sqrt{p}=0.39$.
  The errorbars are results of the numerical simulation in which 
  the size of network is $N=100 \times 100$, and calculated by averaging 
  of $10^5$ samples. 
  The line shows the result from Eq.~$(\ref{eq:energy_mR}).$}
  \label{img:ovlp-engy}
\end{figure}

We investigate the energy landscape around pure retrieval state further.
Let us calculate the energy increase per interaction 
when the state changes from pure retrieval one. 
The energy stored in an interaction is given by
\begin{equation}
 -J_{ij}S_iS_j = -\frac{1}{\sqrt{p}}\sum_{\mu=1}^p 
  \xi_i^{\mu}\xi_j^{\mu} S_i S_j . \label{eq:engy_interaction}
\end{equation}
Substituting $S_i=\xi_i^1,\ S_j=\xi_j^1$ into
Eq.~(\ref{eq:engy_interaction}), 
we obtain the energy stored between $i$-th and $j$-th neurons of the
network retrieving 1-th pattern completely.
As the same way, substituting $S_i=\xi_i^1,\ S_j=-\xi_j^1$ into 
Eq.~(\ref{eq:engy_interaction}) gives the energy of the interaction 
for the case where $j$-th neuron changes from retrieval state.
Subtracting the former from the latter, and averaging it over the 
distribution $(\ref{eq:distribution})$, we obtain the energy increase 
per interaction when the state changes from retrieval one.
\begin{eqnarray}
 [dE_R] &=& 
  \left[ \frac{2}{\sqrt{p}}\sum_{\mu=1}^p
   \xi_i^{\mu}\xi_j^{\mu}\xi_i^1\xi_j^1
 \right] \nonumber\\
 &=&
   \frac{2}{\sqrt{p}}
   \left\{ 
    1+(p-1)C_1\left( \frac{J_0}{\sqrt{p}} \right)
\right\}  .
 \label{eq:engy_increase_R}
\end{eqnarray}
The value of $[dE_R]$ is always positive, and it is expected
that the energy increases on the average when the state changes from pure 
retrieval one. 
We estimate the energies of the finite overlap states as follows. 
Let $n$ be the number of neurons whose states are opposite to the 
retrieval states. 
The relation between $n$ and $m_R$ is given by 
\begin{equation}
 \frac{n}{N} = \frac{1-m_R}{2} \ .
\end{equation} 
If $n$ is not large, the number of interactions between 
the retrieval state and opposite state
is expected to be about $4n$ because the number of nearest 
neighboring neurons is $4$ and neurons whose states are opposite to 
retrieval states lie sparsely almost all case.
Therefore, the energy of the system per neuron whose overlap is $m_R$ is
expected to be 
\begin{eqnarray}
 [E_{m_R}] &=& [E_R] + \frac{4n}{N}[dE_R] \nonumber\\
 &=& [E_R] + 2(1-m_R)[dE_R] . \label{eq:energy_mR}
\end{eqnarray}
Fig.~\ref{img:ovlp-engy} shows the result of numerical simulation 
and compare it with Eq.~$(\ref{eq:energy_mR})$. 
We set parameter values as $p=3$ and $J_0/\sqrt{p}=0.39(>U)$
at the simulation.
We generate $10^5$ sample states for each value of $m_R$ 
and evaluate $[E_{m_R}]$. 
Eq.~$(\ref{eq:energy_mR})$ agrees with the result of the simulation 
near $m_R=1$, although it does not agree with the simulations 
for the case of small $m_R$.
This is clearly because the number of interaction  between 
the retrieval state and opposite state is smaller than $4n$ for large
$n$.
Regardless of it, the energy tends to increase as $m_R$ gets smaller, 
and the state near $m_R=1$ is likely the smallest energy among the 
finite overlap states.
Almost same argument can be applied to the ferromagnetic state, and
the state near $m_F=1$ is the smallest energy  among the ferromagnetic 
order.
Taking into account $[E_R] > [E_F]$, 
the energy of ferromagnetic order is expected to be smaller than that of 
retrieval order
as long as $p \ge 3$ and $J_0 > U\sqrt{p}$ are satisfied.

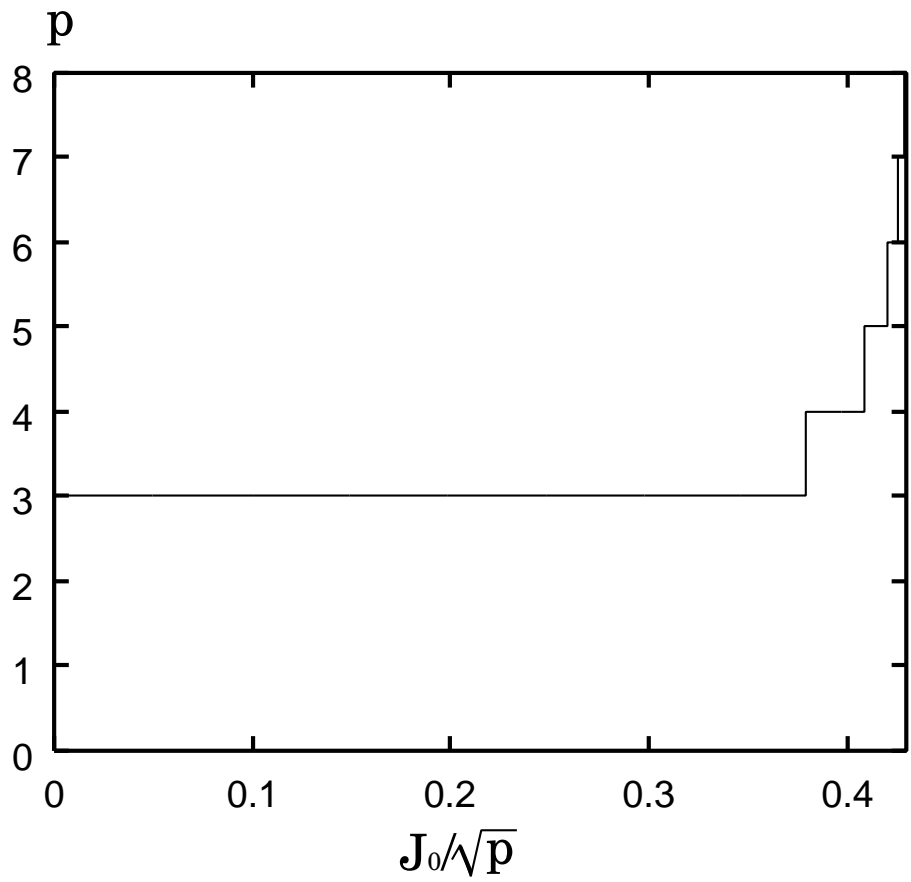
\begin{figure}[htbp]

  \caption{The minimum value of $p$ that is satisfied with 
  Eq.~$(\ref{eq:unstable})$ as a function of the parameter 
  $J_0/\sqrt{p}$ which determines the structure of patterns.}
  \label{img:unstable_p}
\end{figure}

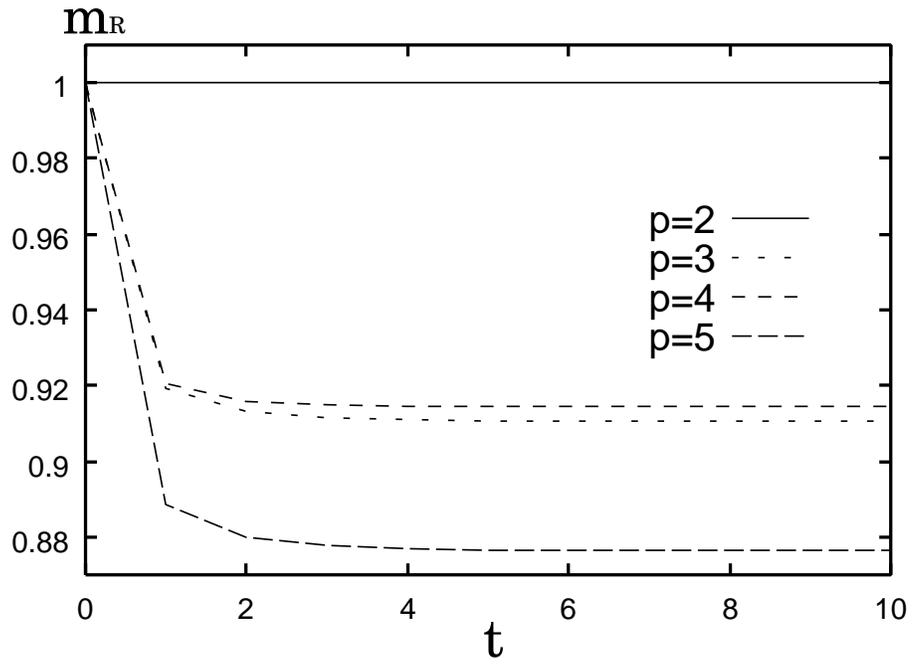
\begin{figure}[htbp]

  \caption{
  The typical example of how the overlap $m_R$ depends on the
  Monte Carlo step $t$ in Monte Carlo simulations at $T=0$.
  We set the system size and the parameter as $N=100 \times 100$ and 
  $J_0/\sqrt{p}=0.43$, respectively.}
  \label{img:unstable}
\end{figure}

We also investigate the stability of pure retrieval state.
The variance of $dE_R$ is given by
\begin{eqnarray}
 [ \{\Delta (dE_R) \}^2] &=& [dE_R^2]-[dE_R]^2 \nonumber\\
 &=& \frac{4(p-1)}{p}
  \left\{ 1+(p-2)C_1\left(\frac{J_0}{\sqrt{p}}\right)^2 
   -(p-1)C_1\left(\frac{J_0}{\sqrt{p}}\right)^4
\right\}  . \label{eq:variance}
\end{eqnarray}
Using Eqs.~$(\ref{eq:energy_R})$ and $(\ref{eq:variance})$, 
The ratio of $[dE_R]^2$ to this variance is obtained by
\begin{equation}
 \frac{[ \{\Delta (dE_R)\}^2]}{[dE_R]^2} =
  \frac
  {(p-1)\left\{ 1+(p-2)C_1\left(\frac{J_0}{\sqrt{p}}\right)^2 
	 -(p-1)C_1\left(\frac{J_0}{\sqrt{p}}\right)^4 \right\}}
  {1+2(p-1)C_1\left(\frac{J_0}{\sqrt{p}}\right)^2+(p-1)^2
  C_1\left(\frac{J_0}{\sqrt{p}}\right)^4} \ .
\end{equation}
Partially differentiating above equation by $p$ 
(the value of $C_1$ is fixed), we obtain the following inequality.
\begin{equation}
 \frac{\partial}{\partial p} 
  \left(
   \frac{[\{\Delta (dE_R)\}^2]}{[dE_R]^2}
  \right) >0  
\end{equation}
This means that the more the value of $p$ gets large, 
the more the energy is likely to decrease when the state changes 
from pure retrieval one, 
although it increases on the average.
Therefore, the pure retrieval state becomes unstable for large $p$.
We can roughly estimate the value of $p$ for which the pure retrieval
state gets unstable by 
\begin{equation}
 \frac{[ \{\Delta (dE_R)\}^2]}{[dE_R]^2} > 1. \label{eq:unstable}
\end{equation}
In Fig.~\ref{img:unstable_p}, we plot the minimum value of $p$ that is 
satisfied with Eq.~$(\ref{eq:unstable})$.
For example, Eq.~$(\ref{eq:unstable})$ is satisfied with $p \ge 8$ 
for $J_0/\sqrt{p}=0.43$. 
Note that this is probably overestimating and the value of $p$ is 
likely smaller.
We confirm this argument in numerical simulations of the dynamics 
at $T=0$.
Fig.~\ref{img:unstable} shows a typical example of how the overlap $m_R$
depends on Monte Carlo step in computer simulations, in which 
$N=100 \times 100$ and $J_0/\sqrt{p}=0.43$.
The state of network starts with the embedded pattern itself.
In this simulations, the pure retrieval state is already unstable 
for $p=3$,
and the network falls into a certain local minimum near $m_R=1$.
This implies that the pure retrieval state is no longer stable
for $p \ge 3$.

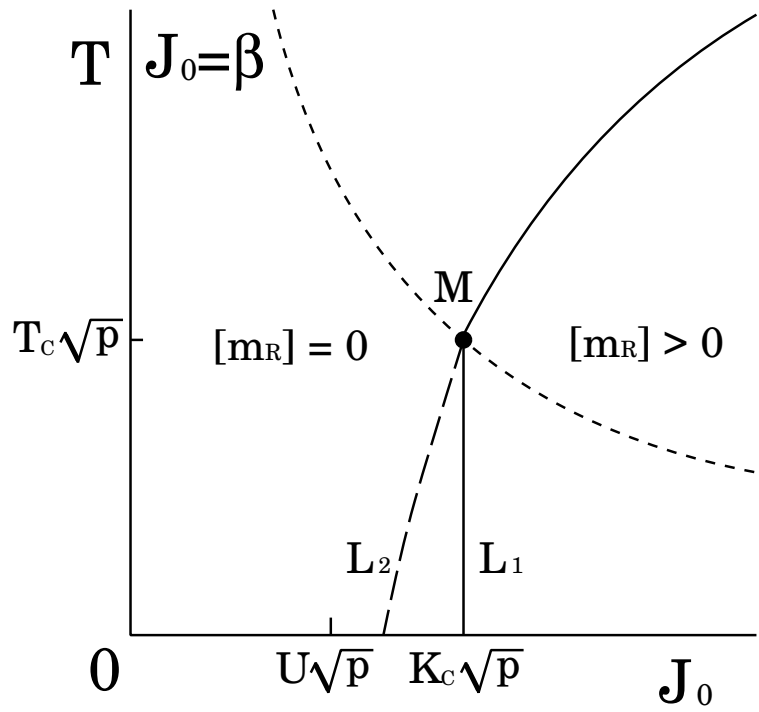
\begin{figure}[htbp]

  \caption{In the case of $p \ge 3$, $U$ is smaller than $K_c\sqrt{p}$ and
  the region of $J_0>U$ at $T=0$ is not retrieval phase.
  Therefore, the boundary of $[m_R]=0$ like $L_2$ is prohibited and 
  it should be like $L_1$.}
  \label{img:proof}
\end{figure}

Taking the above all results into account, 
we show that there is no retrieval phase if $p \ge 3$.
As we have seen, $U$ is always smaller than $K_c=0.44$ for $p \ge 3$,
and the region of $J_0>U$ at $T=0$ is not retrieval phase 
but must be ferromagnetic one.
This is because the ferromagnetic states have smaller energy than 
the retrieval one, and 
the state with minimum energy appears with probability one for $T \to 0$
in terms of the Boltzmann distribution.
Therefore, the boundary of $[m_R]=0$ is prohibited from 
landing at $J_0 < K_c\sqrt{p}$ on the $J_0$ axis like $L_2$ 
in Fig.~\ref{img:proof}, 
and it should land at $J_0 = K_c\sqrt{p}$ from point M 
vertically like $L_1$.
Hence, there is no retrieval phase in the region $J_0<K_c\sqrt{p}$
(for all $T$) 
because the region with finite overlap is limited at 
$J_0 > K_c\sqrt{p}$.
As a result,
the phase diagram should be like Fig.~\ref{img:phase}(b)
and we conclude that the retrieval phase does not exist for the case of 
$p \ge 3$. 

\subsection{The case of $p=2$}
For the case of $p=2$, we cannot immediately conclude that there 
exists the retrieval phase although the energy of a perfect retrieval state 
is smaller than that of ferromagnetic one.
We have to compare the energy of the retrieval state
with that of the spin glass state.
However, it is very hard to calculate the grand state energy of the spin 
glass state because the Hamiltonian of this system has a very 
complicated energy landscape. 
For this case, in order to evaluate the grand state of the Hamiltonian, 
we should carry out simulated annealing, for example.
However, up to now, we do not yet obtain reliable results.
This will be our future problem.

\section{Conclusion} \label{sec:conclusion}
In this paper, we investigated the maximum number of embedded patterns in 
the two-dimensional Hopfield model storing structured patterns.
As a result, we found that 
this system can retrieve a pattern at $T < 2.27$ in the case of $p=1$, 
however the retrieval phase does not exist for the case of $p \ge 3$.
In other words, this system cannot retrieve more than three patterns 
whose value is independent of the size of the system.
This result agrees with a consideration of networks with randomly
diluted synapses, in which $p_c$ is proportional to the average
connectivity per neuron \cite{bib:Derrida}. 
Namely, we can see from this argument that 
the value of $p_c$ for our network is finite 
because the average connectivity per neuron is four 
(finite number).
However, our result is more strict than that from this argument.
  
Moreover, $p_c \le 3$ is satisfied with any value of $J_0$ 
(even the conventional ``random patterns'').
This implies that 
the smallness of the value of $p_c$ is due to the 
dimensionality two of the network rather than spatially correlated 
patterns.
This argument also implies that 
although 
there are several modifications in order to improve storage capacity 
with spatially correlated patterns 
(see \cite{bib:Schluter}, for example), 
these modifications are not expected to be remarkable improvements for our
model system.

From above all arguments,
We conclude that 
storage capacity of associative memory is strongly restricted 
by the spatial structure of the network.

\begin{acknowledgments}
I would like to thank J. Inoue for helpful discussions.
I also thank H. Nishimori for valuable comments.
\end{acknowledgments}



\begin{thebibliography}{6}
\bibitem{bib:Hopfield}J.~J.~Hopfield, Proc. Natl. Acad. Sci. U.S.A. 
	\textbf{79}, 2554 (1982).
\bibitem{bib:Amit}D.~Amit, H.~Gutfreund, and H.~Sompolinsky,
	Phys. Rev. A \textbf{32}, 1007 (1985); Phys. Rev. Lett. \textbf{55},
	1530 (1985).
\bibitem{bib:intro_nc}J.~Hertz, A.~Krogh, and R.~G.~Palmer, 
                    \textit{Introduction to the Theory of Neural Computation}
                      (Addison-Wesley, 1991).
\bibitem{bib:Nishimori}H.~Nishimori, W.~Whyte and D.~Sherrington,
                     Phys. Rev. E \textbf{51}, 3628 (1995).
\bibitem{bib:Onsager}L.~Onsager, Phys. Rev. \textbf{65}, 117 (1944).
\bibitem{bib:Derrida}B.~Derrida, E.~Gardner, and A.~Zippelius, 
	Europhys. Lett. \textbf{4}, 167 (1987).
\bibitem{bib:Schluter}M.~Schl\"uter, F.~Wagner, 
	Phys. Rev. E \textbf{49}, 1690 (1994).
\end{thebibliography}
\end{document}